\documentclass[11pt]{article}
\usepackage[margin=1in]{geometry}
\usepackage[T1]{fontenc}
\usepackage{booktabs}
\usepackage{array}
\usepackage{longtable}
\usepackage{enumitem}
\usepackage[hidelinks]{hyperref}

\newcommand{\muc}{Muon Collider}

\title{\textbf{AI Acceleration of Muon Collider R\&D:\\ A Report from the AI4MuC Workshop}}

\date{}
\author{%
  \parbox{0.95\textwidth}{\centering\normalsize
    \textbf{Authors}\\[0.55em]
    J.~Scott Berg\textsuperscript{1},
    David Bizzozero\textsuperscript{2},
    Tulika Bose\textsuperscript{3},
    Lance Cooley\textsuperscript{4},
    John Dervan\textsuperscript{5},
    Auralee Edelen\textsuperscript{2},
    Peter Elmer\textsuperscript{6},
    Cristiano Fanelli\textsuperscript{7},
    Tova Holmes\textsuperscript{5},
    Walter Hopkins\textsuperscript{8},
    Sergo Jindariani\textsuperscript{9},
    Michael Kagan\textsuperscript{2},
    Maira Khan\textsuperscript{9},
    Young-Kee Kim\textsuperscript{10},
    Matt LeBlanc\textsuperscript{11,12},
    Lawrence Lee\textsuperscript{5},
    Matthias Liepe\textsuperscript{13},
    Radha Mastandrea\textsuperscript{10},
    Patrick Meade\textsuperscript{14},
    Federico Meloni\textsuperscript{15},
    Chad Mitchell\textsuperscript{16},
    Linh Nguyen\textsuperscript{1},
    Isobel Ojalvo\textsuperscript{6},
    Ryan Roberts\textsuperscript{10},
    Nicholas Smith\textsuperscript{9},
    Jean-Luc Vay\textsuperscript{16},
    Gordon Watts\textsuperscript{17},
    Drew Winder\textsuperscript{18},
    Chenran Xu\textsuperscript{8}, and
    Katsuya Yonehara\textsuperscript{9}
  }%
}

\newcommand{\AIEndorserBlock}{%
  \begin{center}
  \begin{minipage}{0.95\textwidth}
    \centering
    \textbf{Endorsers}\\[0.35em]
    \small
    The following individuals endorse this report.\\[0.55em]
       Saptaparna Bhattacharya\textsuperscript{19},
    Martin Kwok\textsuperscript{20},
    Kyoungchul Kong\textsuperscript{21},
    Nate Grieser\textsuperscript{22},
    Ian M. Lewis\textsuperscript{21},
    Zach Marshall\textsuperscript{16},
    Holger Witte\textsuperscript{23},
    Lian-Tao Wang\textsuperscript{10},
    Simone Pagan Griso\textsuperscript{16},
    Yang Ma\textsuperscript{24},
    Javier Duarte\textsuperscript{25},
    Jeffrey Backus\textsuperscript{6},
    Yongcheng Wu\textsuperscript{26},
    Yiheng Ye\textsuperscript{2},
    Karri Folan Di Petrillo\textsuperscript{10},
    Burak Bilki\textsuperscript{27,28},
    Aishik Ghosh\textsuperscript{29},
    Yasar Onel\textsuperscript{28},
    Mark Neubauer\textsuperscript{30},
    Konstantin Matchev\textsuperscript{31},
    Xueying Lu\textsuperscript{32,8},
    Mithlesh Kumar\textsuperscript{1},
    Maria Mazza\textsuperscript{33},
    Georgios Krintiras\textsuperscript{21},
    Maurice Garcia-Sciveres\textsuperscript{16},
    Zhen Liu\textsuperscript{34},
    Regan Doherty\textsuperscript{11},
    Kevin Pedro\textsuperscript{9},
    Ben Simons\textsuperscript{32,9},
    Harvey Newman\textsuperscript{35},
    Mila Bileska\textsuperscript{6},
    Innes Bigaran\textsuperscript{5},
    Carl Vuosalo\textsuperscript{3},
    Adrian Thompson\textsuperscript{36},
    Biao Wang\textsuperscript{28},
    Joe Haley\textsuperscript{37},
    Ryan Schmitz\textsuperscript{38},
    Radovan Dermisek\textsuperscript{39},
    Oz Amram\textsuperscript{9},
    Jaehoon Yu\textsuperscript{40},
    Mary Anne Cummings\textsuperscript{41},
    Yu-Chen Guo\textsuperscript{42},
    Leonid Didukh\textsuperscript{43},
    Emery Nibigira\textsuperscript{5},
    Yuan-Tang Chou\textsuperscript{44},
    Richard Ruiz\textsuperscript{45},
    Gustaaf Brooijmans\textsuperscript{46},
    Mark Larson\textsuperscript{10},
    Abdollah Mohammadi\textsuperscript{3},
    Matthew Herndon\textsuperscript{3},
    Grace Cummings\textsuperscript{9},
    Jullian Watts\textsuperscript{5},
    Jinlong Zhang\textsuperscript{8},
    Jessica N. Howard\textsuperscript{47,48},
    Kiley Kennedy\textsuperscript{49},
    Matheus Hostert\textsuperscript{28},
    Teruki Kamon\textsuperscript{50},
    Aidan Gardner-O'Kearny\textsuperscript{5},
    Marin Mlinarevi\'c\textsuperscript{51},
    Samuel Homiller\textsuperscript{52},
    Maximilian Ruhdorfer\textsuperscript{53},
    Juhi Dutta\textsuperscript{54},
    Muhammad Farooq\textsuperscript{55},
    Reinhard Schwienhorst\textsuperscript{56},
    Toyoko Orimoto\textsuperscript{57}, and
    Chris Rogers\textsuperscript{58}\\[0.7em]
    \footnotesize
 \textsuperscript{1}Brookhaven National Laboratory; 
    \textsuperscript{2}SLAC National Accelerator Laboratory; 
    \textsuperscript{3}University of Wisconsin--Madison; 
    \textsuperscript{4}National High Magnetic Field Laboratory; 
    \textsuperscript{5}University of Tennessee, Knoxville; 
    \textsuperscript{6}Princeton University; 
    \textsuperscript{7}William \& Mary; 
    \textsuperscript{8}Argonne National Laboratory; 
    \textsuperscript{9}Fermi National Accelerator Laboratory; 
    \textsuperscript{10}University of Chicago; 
    \textsuperscript{11}Brown University; 
    \textsuperscript{12}The NSF AI Institute for Artificial Intelligence and Fundamental Interactions (IAIFI); 
    \textsuperscript{13}Cornell University; 
    \textsuperscript{14}Stony Brook University; 
    \textsuperscript{15}Deutsches Elektronen-Synchrotron DESY; 
    \textsuperscript{16}Lawrence Berkeley National Laboratory; 
    \textsuperscript{17}University of Washington; 
    \textsuperscript{18}Oak Ridge National Laboratory; 
    \textsuperscript{19}Southern Methodist University; 
    \textsuperscript{20}University of Nebraska--Lincoln; 
    \textsuperscript{21}University of Kansas; 
    \textsuperscript{22}University of Cincinnati; 
    \textsuperscript{23}Massachusetts Institute of Technology; 
    \textsuperscript{24}UCLouvain; 
    \textsuperscript{25}University of California, San Diego; 
    \textsuperscript{26}Nanjing Normal University; 
    \textsuperscript{27}Turkish Accelerator and Radiation Laboratory; 
    \textsuperscript{28}University of Iowa; 
    \textsuperscript{29}Georgia Institute of Technology; 
    \textsuperscript{30}University of Illinois Urbana-Champaign; 
    \textsuperscript{31}University of Alabama; 
    \textsuperscript{32}Northern Illinois University; 
    \textsuperscript{33}Florida State University; 
    \textsuperscript{34}University of Minnesota; 
    \textsuperscript{35}California Institute of Technology; 
    \textsuperscript{36}University of California, Irvine; 
    \textsuperscript{37}Oklahoma State University; 
    \textsuperscript{38}Imperial College London; 
    \textsuperscript{39}Indiana University; 
    \textsuperscript{40}University of Texas at Arlington; 
    \textsuperscript{41}Muons, Inc.; 
    \textsuperscript{42}Liaoning Normal University; 
    \textsuperscript{43}SNCT; 
    \textsuperscript{44}National Tsing Hua University; 
    \textsuperscript{45}Institute of Nuclear Physics, Polish Academy of Sciences; 
    \textsuperscript{46}Columbia University; 
    \textsuperscript{47}University of California, Santa Barbara; 
    \textsuperscript{48}Kavli Institute for Theoretical Physics; 
    \textsuperscript{49}Tufts University; 
    \textsuperscript{50}Texas A\&M University; 
    \textsuperscript{51}CEA Paris-Saclay; 
    \textsuperscript{52}University of Pittsburgh; 
    \textsuperscript{53}Stanford University; 
    \textsuperscript{54}Department of Physics, School of Advanced Sciences, Vellore Institute of Technology, Vellore, India; 
    \textsuperscript{55}University of New Hampshire, Durham, New Hampshire, United States; 
    \textsuperscript{56}Michigan State University; 
    \textsuperscript{57}Northeastern University; 
    \textsuperscript{58}STFC Rutherford Appleton Laboratory
  \end{minipage}
  \end{center}
}

\begin{document}
\maketitle

\AIEndorserBlock
\vspace{2em}
\begin{abstract}
\noindent
A \muc{} is one of the most attractive routes to multi-TeV lepton collisions, but it requires the accelerator and detector communities to solve a tightly coupled set of problems that have no close precedent in an operating collider. Muons must be produced, captured, cooled, accelerated, and collided before they decay. The detector must then extract clean physics from an environment dominated by decay products from the beamline. Artificial intelligence is not a substitute for accelerator physics, detector design, or engineering validation. It is, however, unusually well matched to several of the central obstacles: high-dimensional design spaces, expensive simulations, coupled subsystem trade-offs, and workflows that are currently too slow or too fragile to support broad design exploration. This report distills the AI4MuC workshop into a summary of AI use cases with the potential to change the pace of \muc{} R\&D, organized around the following physics bottlenecks: muon production and capture, ionization cooling, magnet and RF technology, detector design in beam-induced background, background simulation and generation, reconstruction, and software. 
\end{abstract}

\vspace{1em}

\noindent
The 2023 P5 report identified a \muc{} as the ``muon shot'' for the future of US particle physics, and the National Academies report on elementary particle physics subsequently made it the highest recommendation for the long-term vision of the field~\cite{P5:2023wyd, NationalAcademiesofSciencesEngineeringandMedicine:2025bix}. The community that must deliver this machine is growing---the US Muon Collider Collaboration (USMCC) now counts nearly three hundred members---but it remains small compared to the scale of the design problem it has taken on. That mismatch is the context for this report.

The \muc{} design problem is a complex, heterogeneous chain of tightly coupled elements, each expensive to simulate. Changes to these configurations, within realistic engineering constraints, have a profound impact on the downstream physics, including in the optimal configurations of downstream components. Independent iteration of component design will fail to find an optimal solution, and iterating across all possible configurations is unfeasible. Moreover, the metrics are neither simple nor uniform: one system may be heavily dependent on tail modeling while another is focused on losses through beam decays. Most systems have multiple competing objectives, like the interaction region's need for both high luminosity and low levels of beam-induced backgrounds. Producing a facility-wide optimization will be challenging. 

The AI4MuC workshop \cite{AI4MuC2026}, held June 29--30, 2026, brought together accelerator physicists, detector experts, AI/ML practitioners, and members of adjacent programs to ask where AI and ML can improve a \muc{} design loop, and what evidence would make the result trustworthy. The workshop aimed to integrate AI/ML expertise into the existing US \muc{} effort and to identify concrete applications for AI/ML tools. One theme recurred throughout: the most valuable advantage from these tools is to leverage limited resources. Well-validated automation can turn sparse expert time into tangible progress, expanding the reach of the R\&D effort.

The goal is to use AI/ML to make better \muc{} decisions: to identify robust target and capture designs, improve cooling channels, explore magnet and RF concepts under coupled constraints, map detector trade-offs in the beam-induced background (BIB) environment, generate validated fast BIB samples, and make software workflows reproducible enough that many groups can contribute. Near-term deliverables should be validated design loops, not just a catalog of algorithms. We recommend the community organize its AI/ML efforts around a small number of benchmark problems with agreed-upon physics metrics, trusted validation samples, and public workflows, each connected to a central challenge of \muc{} R\&D.

\section{The Promise and Challenge of a Muon Collider}

A \muc{} is attractive because it could deliver elementary-particle collisions at multi-TeV energies in a compact circular machine; it is difficult because the particle that makes this possible decays on the timescale in which the machine must achieve the desired luminosity.

This fact propagates through the entire facility design. An intense, short-pulsed proton driver must interact with a resilient target to produce enough pions to begin the process. The front end must capture particles within a large phase space acceptance. The cooling channel must reduce the six-dimensional emittance of the resulting muon beam by roughly six orders of magnitude, in the presence of transmission losses and muon decay. Acceleration must be rapid. The collider ring must handle intense beams whose decay products create radiation, heat loads, and detector backgrounds. The detector must reconstruct physics objects in the presence of large backgrounds unlike those of either a hadron collider or an electron--positron collider.

Ionization cooling is often considered the flagship technical challenge of a \muc{}, but the underlying physics of energy loss and re-acceleration is straightforward, and its application has been demonstrated in both stand-alone experiments and operational accelerators \cite{MICE:2019jkl, Bradley:IPAC2018-TUPMK015}. The difficulty is in scaling: addressing the engineering challenges that come with extending this principle to the high-intensity beams of a \muc{}, and building cells capable of extending the iterative cooling process much further than what's been achieved before. This challenge incorporates a large number of critical \muc{} technologies, including high-field magnets, high-gradient RF, and robust targets and absorbers, all operating in close proximity to one another.
Designs must be optimized considering the constraints of and interrelations between each of these systems, which is precisely the regime in which modern AI methods have shown promise in other fields.

The workshop agenda reflected the diversity of challenges: overview and community organization talks were followed by sessions on differentiable tracking, accelerator tuning, targetry, magnets, RF, ionization cooling, detector design, detector R\&D, BIB simulation, generative BIB, software and reconstruction, agents, and lessons from the Electron-Ion Collider (EIC) AI program.

\section{A snapshot of today's AI integration}

The diverse fields that are part of \muc{} R\&D are each interacting with AI/ML in different ways, creating an uneven starting point but a great deal of opportunity for inter-community sharing.

The accelerator community already has substantial experience with Bayesian optimization, reinforcement learning, differentiable tracking, surrogate modeling, diagnostics, online tuning, and anomaly detection. AI-assisted tuning workflows at operating facilities have reduced tasks that took hours of operator work to a few minutes, and digital-twin prototypes are running at multiple accelerators, with model updates driven by live machine data. These methods are most effective when objectives and constraints are explicit. For a \muc{}, however, the most important accelerator questions are rarely single-objective. A cooling channel is judged not only by an emittance metric but by which particles it successfully transports. A target is judged not only by its yield but by whether it survives its own stress and heat loads. A magnet surrogate is judged not only by its smooth field but by whether its stability limit is predictable. Optimization should therefore use multi-objective, constraint-aware metrics constructed considering the broader facility performance.

On the detector side, modern high-energy physics already uses machine learning extensively for reconstruction, object identification, triggering, fast simulation, and analysis. Nonetheless, the \muc{} BIB environment differs from LHC pileup in several respects that make its handling distinct. It is diffuse, out-of-time, and enters the detector region primarily from the nozzles, distinctly separating it from collision products in its features. Bunch crossings at a \muc{} are expected to occur at roughly one-thousandth the rate of the LHC, making read-out and triggering requirements fundamentally different. While these differences both point to systems designed to filter out any non-pointing objects, that background rejection must be balanced against detector sensitivity to unconventional signatures of new physics. Finally, the simulation of the BIB is lacking a dataset by which to validate it, making detector design robustness to variations in BIB essential.


On the software and computation side, the opportunity is more immediate. The design program is fundamentally limited by the computational cost of simulation. Surrogate models are being developed to ease some of the computational cost of physics studies, but production of training data and a full understanding of tails cannot be shortcut by this method. Reproducible workflows with built-in testing are beginning to appear, but more work is needed. This includes interconnect with the machine-detector interface simulation, as well as detector geometry, simulation, digitization, reconstruction, validation, and analysis. LLM agents and retrieval-based tools can help if they are constrained to auditable workflows, and can help build automated testing suites. They are not reliable enough to replace experts, but they are already useful in conjunction with them: designing streamlined tools and building connections between disparate software that improve the usability of the \muc{} simulation chain.

\muc{} R\&D can benefit from a wide landscape of AI tools. Differentiable simulators, surrogate models and Bayesian optimization can help reduce expensive scans to manageable sizes. Generative models can help produce background simulation, while foundation models can be considered for reconstruction strategies. Anomaly detection is already being used across targetry and magnet development. Each unique \muc{} challenge comes with a suite of possible AI/ML solutions.

These applications can also contribute to AI research. \muc{} problems can stress-test models on high-dimensional and heterogeneous scientific data with rare tails and shifting distributions. Workflows will require a mixture of automation and expert intervention. Public benchmarks, transferable models, and validation protocols built for this effort have the potential to transfer back to the broader AI community. 

\section{A challenge map}
\label{sec:map}


{\small
\begin{longtable}{>{\raggedright\arraybackslash}p{2.4cm}>{\raggedright\arraybackslash}p{4.5cm}>{\raggedright\arraybackslash}p{3.6cm}>{\raggedright\arraybackslash}p{4.2cm}}
\toprule
\textbf{Challenge} & \textbf{Why it is hard} & \textbf{Where AI can plausibly help} & \textbf{What would count as success} \\
\midrule
Muon production and capture &
Design of the MW-class targetry requires pion yield and initial muon emittance optimization, while understanding the material response, shielding requirements, capture magnet acceptance, radiation effects, and maintainability. &
Surrogates for expensive material simulations; multidimensional optimization of all elements; incorporating operational experience from existing facilities. &
A model that gives rapid optimization in the $\pi$ yield, emittance, material stress, and acceptance space that agrees with trusted simulations. \\
\addlinespace
Muon ionization cooling &
6D emittance must fall by six orders of magnitude in a channel containing absorbers, RF, high-field solenoids; requires modeling of beam dynamics and decay, stochastic material effects, response of candidate absorber materials, RF in high $B$ fields, magnet engineering. &
Differentiable tracking, particularly in presence of stochastic effects; gradient-guided lattice optimization; surrogate models for staged channels; uncertainty-aware tuning. &
Improved cooling designs validated against trusted stochastic tracking for phase space distributions including tail effects, transmission rates, and sensitivity to material effects. \\
\addlinespace
Magnets &
Frontier-class high-field solenoids required; fast-ramping magnet technology directly affects accelerator complex design; Field quality, aperture, stress, quench margin, conductor limits, cryogenics, radiation, safety, cost, and manufacturability are coupled. &
Fast multiphysics surrogates; quench-margin prediction; operational anomaly detection; design-space exploration. &
Multi-input AI-driven quench detection and handling; multi-dimensional mechanical models aid design optimization. \\
\addlinespace
RF technology &
Cavities must satisfy electromagnetic, thermal, mechanical, breakdown, integration, and lattice constraints, often in strong magnetic fields. &
Bayesian or inverse design; agent-assisted CAD-to-simulation loops; multiphysics surrogate models. &
Reproducible, expert-verifiable field, power, thermal, stress models. \\
\addlinespace
Detector design in BIB &
Large muon decay background in detector constrains geometry, shielding, timing, segmentation, reconstruction requirements and physics reach.  &
Cross-detector-subsystem reconstruction optimization; surrogate detector response; AI-guided Pareto-front mapping for reconstruction and shielding design, automated calibration. &
Detector choices compared using agreed physics metrics: efficiency, fake rates, resolution, timing rejection, displaced-object performance, and robustness to BIB variations. \\
\addlinespace
Smart detector R\&D &
Local rejection of BIB-like hits must obey power, latency, area, mass, and radiation constraints while preserving physics signals. &
On-detector ML; spiking networks; interpretable local classifiers. &
Demonstrated BIB rejection at fixed efficiency for prompt and nonprompt signals, with latency/power estimates and rare-tail tests. \\
\addlinespace
BIB simulation and generation &
Detailed BIB simulation poses major computational challenge; Full samples are costly; simplified samples may miss correlations and rare high-occupancy tails. &
Conditional generative models (e.g. normalizing flows, diffusion models); optimization of full sim workflow. &
Generative model agreement with full sim in spectra and correlations; Significant reduction in full sim computation needs. \\
\addlinespace
Software and reconstruction workflows &
Design studies are often one-off and difficult for non-experts to reproduce, compounded by immense computational requirements. &
LLM agents with restricted tools; documentation interpretation; automated validation reports; automated containerized workflows. &
Agentic workflow execution; agentic development loops; automated output validation. \\
\bottomrule
\end{longtable}}

\section{Muon production and capture}


The target system must convert the high-intensity incident proton beam into pions, the capture system must collect useful phase space, and all of the hardware involved must survive the resulting radiation. The design objective must balance pion production, captured muon phase space, heat deposition, shock, radiation damage, shielding, remote handling, activation, and downstream cooling acceptance.

The workshop discussion identified simulation benchmarking against experiment as the limiting factor for progress in this area. Radiation damage and material response at the relevant beam parameters are not well covered by existing databases---much of the materials-science literature, developed for nuclear engineering, addresses different energy scales---so reliable simulation requires dedicated benchmarking experiments. Target-relevant data are also scattered across institutions in incompatible formats. An early, high-value step is an integrated database of target materials and irradiation data.

This is also the area where the \muc{} effort has the most to gain from its neighbors. Operating and planned high-power target facilities---the SNS, LBNF/DUNE, Mu2e, and J-PARC---face closely related problems, and the RaDIATE collaboration already coordinates radiation-damage studies of beam-intercepting materials. The SNS experience presented at the workshop showed both what is possible (AI-assisted analysis of target strain and lifetime data at an operating megawatt facility) and what is painful (heterogeneous, partly Windows-bound engineering software that resists being placed inside an automated optimization loop). The natural strategy is to develop common tools and workflows with these programs rather than duplicating them, and several university groups---including students already working on the \muc{} demonstrator target design---are well-poised to contribute. As with all optimization problems, care is needed to prevent problems such as over-ambitious target designs that cause later problems due to short lifetimes or unreliable operation. 

AI can be valuable here because the target/capture problem is multi-objective and expensive. A brute-force scan over target material, geometry, beam parameters, capture-field configurations, shielding assumptions, and cooling acceptance is unlikely to be sufficient. A physics-informed surrogate could learn a local map from design variables to yield and engineering-relevant observables. Active learning could decide which expensive simulations or facility measurements are most informative. Anomaly detection and lifetime modeling could connect the \muc{} target program to operating high-power facilities.

The model output passed downstream must include accepted phase space, correlations, time structure, and loss patterns rather than only a pion or muon count. A near-term target benchmark should therefore couple a target/capture simulation to a downstream acceptance metric and explicitly define the information required by the cooling design.

\section{Ionization cooling}


Ionization cooling is a unique accelerator design challenge for a \muc{}. The beam emerging from the front end is too large in phase space for a high-luminosity collider by roughly six orders of magnitude in six-dimensional emittance. The mechanism for closing that gap is by increments: a well-designed cooling cell reduces the emittance by roughly ten percent, so the kilometer-scale rectilinear cooling channel design~\cite{Stratakis:2014nna} strings together on the order of a hundred cells to achieve the required cooling. The later stages require high-field magnets with robust radiation tolerance operating with RF in close quarters, and simulating the channel requires accurate modeling of absorbers, RF cavities, solenoids and wedge materials, considering the effects of dispersion, multiple scattering, energy straggling, longitudinal dynamics, aperture, and stage-to-stage matching.

The dimensionality of the parameter space makes manual optimization difficult. Lattice designs require specification of RF gradients, magnet strengths, absorber geometries, and in some cases, carefully calibrated angles or compound magnets~\cite{InternationalMuonCollider:2024jyv,Fabbri:2025cfg,Riggall:2026fju}. A recent IMCC redesign of the rectilinear 6D and final cooling channels, using conventional optimization, reported a factor-of-two reduction in final transverse emittance relative to the MAP-era designs, reaching the design target for the first time \cite{zhu:ipac2026-wep4655}. If this simpler re-optimization results in a factor of two improvement, a systematic, gradient-informed exploration of the full parameter space has strong potential to produce even better performance.

Differentiable simulation offers a way to identify high-dimensional directions that hand scans and black-box optimization may miss. Applied to a cooling channel, gradients can expose influential parameters, dangerous tolerances, and combinations of local changes that improve luminosity. This is an area of active development, and one report at the workshop described progress in building a ground-up differentiable cooling channel and validating its first elements against IMCC benchmarks \cite{Smith:beamline}. The outstanding challenge is extending this process to emulate the stochastic processes of absorber interactions. 

Care must be taken in accounting for the multiple physical processes as well as in defining a sound objective. Cooling is not just smooth optics: stochastic absorber effects, tails, losses, scraping, aperture, and transmission are central to the answer, and a model that differentiates an idealized map but fails on straggling or multiple scattering will be misleading. On the objective, emittance is not sharply defined in a non-conservative system, and a sensible starting point is to work with eigenvalues of the full six-dimensional covariance matrix under a Gaussian assumption while the community converges on better figures of merit. The correct benchmark is distribution-level agreement with trusted tracking, including emittance, transmission, acceptance, tails, and sensitivity to material and lattice perturbations. If the differentiable model is approximate, the approximation should be explicit and its domain of validity should be documented.

The workshop close-out identified a structural next step: connect the groups that have separately optimized individual cooling systems, because global optimization---from capture through initial cooling and across stage-to-stage matching---remains open. A first benchmark could use a restricted section of the rectilinear channel with a small number of tunable elements, and compare manual, black-box, differentiable, and surrogate-assisted optimization using the same initial distribution and the same validation plots. Deliverables should include phase-space projections, emittance evolution, losses, transmission, and parameter sensitivities in addition to model-level scores.

\section{Magnets and RF}

The \muc{} design is significantly constrained by technology development. High-field capture solenoids, focusing solenoids for cooling, rapid acceleration systems, collider-ring magnets, and RF cavities all sit near difficult boundaries. These systems are not independent components. A magnet design changes aperture, optics, shielding, forces, cryogenics, and integration. An RF design changes acceleration, heating, breakdown risk, lattice geometry, and compatibility with nearby magnetic fields \cite{Boscolo2019MuonCollider}.

For the high-field magnets, more data is needed to fully understand today's constraints. Quench behavior is reasonably characterized for low-temperature superconductors, but the corresponding datasets for high-temperature superconductors---on which the most ambitious \muc{} solenoids depend---largely do not yet exist, and even with dedicated test programs the full critical surface cannot be explored experimentally. Efficient, physics-informed simulation and surrogate modeling are needed to extend predictions into regions that cannot be measured directly \cite{Marchevsky2021HTSQuench}. The fusion industry is currently driving high-field magnet development, but much of the resulting data is proprietary, and the national laboratories are actively working on access and on open datasets---an area where community-level coordination can have impact.

This makes magnets and RF good candidates for AI-assisted multiphysics design, but poor candidates for unconstrained black-box optimization. The useful role for AI is to shorten expert iteration cycles while preserving interpretable outputs. For magnets, that may mean surrogates for field quality, stress, stability margin, or quench-relevant observables, trained on simulation and test-stand data with calibrated uncertainties \cite{Bottura2025MagnetRD}. For RF, it may mean restricted inverse-design or Bayesian optimization workflows that automatically produce the plots experts already use: field maps, gradients, power deposition, thermal deformation, stress, and breakdown-related indicators.

Agentic workflows are particularly useful here. Much of the time in multiphysics design is spent moving between CAD, meshing, electromagnetic simulation, thermal and mechanical analysis, plotting, and documentation. SLAC is already operating an internal LLM coding service with RF and accelerator applications, and hosts regular internal meetings on practice and institutional IT support, including a deliberate effort to package validated agent ``skills'' so that a workflow developed for one accelerator application can be transferred to another. A well-constrained agent that can create a versioned simulation deck, run a standard chain, parse logs, make validation plots, and flag failures can be useful without being asked to make final engineering judgments. The benchmark should be reproducibility and expert auditability: can the workflow produce the same plots, with the same inputs, and can a human understand why a candidate design passed or failed?

\section{Detector design with intense backgrounds}


The beam-induced background (BIB) intrinsic to a \muc{} is the driving factor for detector design. 
Muon decays along the beamline generate particles that enter the detector with distinct spatial, timing, and energy structure. 
The first-order defense is massive tungsten shielding: nozzles in the forward regions absorb about 99.9\% of the shower energy from each TeV-scale decay electron. 
Nonetheless, with millions of decays contributing per bunch crossing, even the small escaping fraction produces a substantial diffuse, soft, and out-of-time background throughout the detector. 
Shielding and nozzle geometry, timing layers, tracker layout, calorimeter segmentation, reconstruction thresholds, and trigger or readout strategies are therefore coupled design variables.

Evaluating the physics performance of a single detector configuration requires thousands of CPU hours and substantial expert attention, making broad design scans prohibitively expensive. AI can help quantify the coupled trade-offs. A detector optimization campaign should map a Pareto front across physics efficiency, fake rate, energy and momentum resolution, timing rejection, forward acceptance, displaced-object performance, bandwidth, cost, and robustness to BIB variations. This representation lets designers compare explicit trade-offs rather than only the small number of configurations affordable with a full simulation.

The EIC experience presented at the workshop is directly relevant because it treats detector design as exactly this kind of multi-objective problem, with physics benchmarks, reproducible optimization pipelines, and now LLM-assisted workflow monitoring with humans in the loop, in a collaboration of over a thousand members \cite{diefenthaler2024ai,anderson2026scalable}. 
%
It also addressed a common objection: detector design and reconstruction are coupled, but this dependence should remain manageable. Many design variations do not require changes to the reconstruction, while reconstruction strategies with reduced dependence on detector-specific assumptions can further limit configuration-specific retuning. Where explicit dependence remains, design parameters can be propagated consistently through simulation and reconstruction within a fiducial region of the design space.

The most important detector-design benchmarks should be tied to physics performance. Representative benchmarks include high-momentum muons, electrons and photons, jets, missing momentum, heavy-flavor signatures, displaced tracks or vertices, and high-mass final states. Each should be evaluated with and without realistic BIB overlays and with systematic variations in the background model. 
%
The relevant question is not whether an ML method performs well on an intermediate reconstruction task, but whether the resulting detector and reconstruction strategy delivers the physics performance required across the target benchmark channels.

\section{On-detector intelligence}


The large, diffuse BIB activity in a \muc{} detector creates a strong motivation for implementing machine-learning algorithms directly in the detector readout hardware, where online background rejection could significantly reduce the volume of data that must be read out. At the same time, the relatively long interval between bunch crossings may permit the use of more complex algorithms with longer latencies or greater power consumption than would be practical at the LHC. The vertex detector experiences the densest background, with occupancies projected to lie in the range $1-10\%$ and data rates of up to $75\ \mathrm{Gbps/cm^2}$ \cite{VertexDataRate}. These conditions exceed those expected for the Phase-2 pixel detectors of the ATLAS and CMS experiments at the HL-LHC.


Explorations of small, quantized neural networks (NNs) for on-detector BIB suppression in the pixel detector, building on the work of the Smart Pixels Collaboration \cite{SmartPixFiltering,SmartPixRegression}, have already demonstrated promising performance using a simplified simulation of the background. A number of architectures have been explored, ranging from simple networks using basic cluster sizes and locations as inputs to convolutional neural networks (CNNs) operating on entire pixel clusters. The tradeoff between background rejection and hardware resource usage has been studied by synthesizing FPGA implementations of these networks as a proxy for their relative hardware complexity. These studies identified a model of comparable size to that of Ref.~\cite{SmartPixFiltering} that rejects approximately 85\% of background hits while retaining 99\% of signal hits. Additional studies have investigated the impact of single-event upsets (SEUs) caused by radiation-induced bit flips in the chip’s digital logic. Early results indicate that the NN can be sensitive to bit flips in its weights, but also suggest that targeted mitigation strategies may be more efficient than triplicating the full NN logic.


Significant work remains to establish whether such algorithms can be incorporated into a realistic detector ASIC. This includes developing the logic required to identify clusters, pipeline data into the NN, and set and tune multiple thresholds while satisfying constraints on latency, power, area, and data throughput. The algorithm and ASIC implementation must also remain robust against changing detector conditions, particularly radiation damage to the sensor and changes in operating bias voltage. Equally importantly, background rejection must maintain high signal efficiency across the broad range of physics signatures accessible at a \muc{}, including high-occupancy jets, displaced decays, and slow-moving particles. In parallel with more detailed studies of hardware implementation and robustness, a broader range of approaches is being investigated, including neuromorphic architectures and logic-gate networks.

\section{BIB simulation and generative models}

BIB is produced when muon decay products from the beam interact with the magnetic fields and materials of the \muc{} accelerator lattice and detector. Accurate BIB simulation is a necessary tool for optimizing detector layout, readout, reconstruction, physics sensitivity, and smart detector R\&D. 

During the workshop's detector close-out session, generating BIB simulation was identified as the current chief bottleneck of the detector program. This simulation involves tracking the decay products of the circulating beams through up to two hundred meters of shielding and machine elements, and includes a spectrum of physics processes across different energy scales and particle species. The compute power required is massive: producing even a single event consumes weeks of CPU time and tens of GB of disk space. 

Because of these limitations, current BIB simulation takes a FLUKA sample representing 10\% of a single bunch crossing's parent muon population, clones and randomizes each parent's decay products in azimuth to inflate the apparent statistics, and then passes the resulting pseudo-events through full detector simulation. This approach is functional but has known physical costs. Azimuthal cloning washes out the genuine asymmetries imprinted by the machine lattice: BIB is dominated by decays inside the final focus quadrupoles and chicane dipoles, which produce particles with specific directional signatures that symmetrization erases. The finite reused sample also propagates its structural artifacts, including the statistical underrepresentation of rare decay events, into every overlaid physics event. These rare events are physically important: secondary Bethe--Heitler muons, produced in photonuclear reactions and concentrated at specific production hotspots, remain poorly characterized precisely because the current sample lacks the statistics to study their tails. A serious campaign of nozzle or lattice optimization, where design changes necessarily invalidate previously simulated background, requires this pipeline to become significantly cheaper.  Complementary efforts to restructure the full simulation itself are underway, and it is likely that these efforts will make use of fast machine learning generative models. 

The workshop saw a first attempt at developing a generative machine learning model for BIB simulation using conditional normalizing flows (neural spline flows adapted to the cylindrical detector geometry) and tabular diffusion models \cite{mastandrea2026fastbibsimulationfuture}. These models were trained to generate pre-digitization tracker BIB hits in five kinematic observables (energy, time, and $xyz$ position), conditioned on the tracking detectors' structure (i.e. \textit{side} and \textit{layer} indices). A separate model was trained for each tracking detector subsystem. The marginal distributions for the five kinematic observables were reproduced well across subsystems after geometric post-processing. In addition, post-digitized and reconstructed tracks generated from the generative model's BIB hits showed good agreement with those generated from the full simulation BIB training set. 

The present implementations generate hits independently and therefore cannot represent hit-to-hit or event-level correlations, though this could be improved with model changes and a more comprehensive dataset. In particular, the current dataset's azimuthally cloned sample makes event-level correlations impossible to learn. 
Other ongoing projects on the topic of BIB simulation involve simulating pre-digitization calorimeter hits and using more sophisticated diffusion models to improve upon the conditional normalizing flow results.

To proceed, the community should define a validation strategy for generative BIB including:
\begin{enumerate}[itemsep=1pt]
\item \textbf{Marginal distributions:} particle species, energy or momentum, timing, radius, longitudinal position, angle, and detector-layer occupancies.
\item \textbf{Correlations:} layer-to-layer occupancy, time--position structure, regional correlations, event-level fluctuations, and high-occupancy tails.
\item \textbf{Detector consequences:} cluster merging, fake tracks, timing rejection, calorimeter response, missing momentum, object reconstruction, and signal efficiency.
\item \textbf{Design decisions:} whether the same detector or reconstruction choice would be made using full simulation and the generative sample within agreed uncertainties.
\end{enumerate}

\section{Reconstruction, software, and agents}
\label{sec:software}

The software challenge for a \muc{} extends beyond the improvement of reconstruction algorithms: the design program of the entire facility itself must be rapidly reproducible.
A detector optimization result is only useful if the geometry, samples, digitization, reconstruction, software release, configuration files and validation plots that produced it are accessible and understandable to downstream researchers.
Reproducibility in this context determines whether a design study becomes a shared asset of the \muc{} community, or remains a private result.

There has been substantial progress on the \muc{} software stack, which was recently restructured for the 3.x release series, the first tag of which was issued in June 2026 shortly before the AI4MuC workshop.
The legacy Marlin/ILCSoft~\cite{Gaede:2006pj,ilcsoft} reconstruction chain was retired in favor of a Gaudi-native workflow~\cite{Barrand:2001ny} built on the Key4hep ecosystem~\cite{Ganis:2021vgv}, with EDM4hep~\cite{Gaede:2022leb} as the default event data model.
The ACTS tracking toolkit~\cite{Ai:2021ghi} has been integrated, and reconstruction algorithms developed for the \muc{} are contributed to the shared Key4hep repositories. Additionally, ML-based particle-flow (MLPF) reconstruction, which has been applied in CMS at the LHC~\cite{CMS:2026znb}, CLICdet at CLIC~\cite{Pata:2023rhh}, and CLD at FCC-ee~\cite{Mokhtar:2025zqs}, is being adapted for the MAIA \muc{} detector concept.

Releases are assembled with Spack~\cite{Gamblin:2015hhi} and distributed as layered container images, comprising an analysis layer that provides EDM4hep/podio and AI/ML tooling, and an optional simulation layer adding reconstruction, \textsc{Geant4}~\cite{AGOSTINELLI2003250} and the \textsc{Whizard}~\cite{Kilian:2007gr}, \textsc{MadGraph5\_aMC@NLO}~\cite{Alwall:2014hca} and \textsc{Pythia}~\cite{Bierlich:2022pfr} event generators.
These images are built for both x86\_64 and aarch64 and replicated to \verb|CVMFS|~\cite{Blomer:2011zz}.
Three detector concepts are selected through a single configuration switch that is applied to a common pipeline running through generation, simulation, BIB overlay \& digitization, reconstruction and analysis.

Of particular relevance in the context of the AI4MuC workshop was the automated physics validation that is now running in continuous integration.
Each commit to the \verb|mucoll-spack| main branch triggers a per-particle simulation, digitization, reconstruction and plotting chain that is executed against the newly built image.
The resulting performance plots are published for each of the validated object types (muons, electrons, charged pions and photons) and labeled with the analysis settings that were used to produce them.
The software stack is open source~\cite{mucollsoft} and is accompanied by a maintained tutorial~\cite{mucolltutorial}; the community is actively seeking additional contributors.

This infrastructure is a precondition for the productive use of LLM agents. The
utility of an agent is bounded by the environment it is able to enter, by
whether failures are reported rather than silently absorbed, and by the
availability of reference output against which its results can be checked. A
versioned container, a well-documented configuration, and continuously regenerated
reference plots supply all three.
Recent results from outside the \muc{} community indicate that these conditions
are close to sufficient for successful agent workflow integration. On a benchmark of eight analysis tasks, each requiring
the production of a specified physics plot from a specified dataset using the
community software stack, frontier-model agents operating in a simple
execute--inspect--retry loop achieved exact-success rates above 80\%, with
clear cost--performance trade-offs across models~\cite{Watts2026LLMEcosystem}.
Complete library implementations can also be generated unattended from a
sufficiently detailed specification: a four-page specification of a
computational-graph library for full-systematics analysis was supplied to a
coding agent, which produced a minimal, tested, and functional library at a
compute cost of order one hundred dollars, and following expert review it is
expected to enter the working analysis stack~\cite{Gray2026AgenticEngineering}.
A first end-to-end agentic physics analysis has likewise been
demonstrated~\cite{Watts2026LLMEcosystem}. The underlying capability is
therefore established; what remains is an agreed means of measuring it against
\muc{} workflows.

A suitable benchmark can be constructed from the validation infrastructure
already in place. Given a tagged release, a geometry configuration, and a
versioned reference sample, an agent would be required to produce a standard
validation report: the per-particle performance plots generated by the current
continuous-integration chain, together with the command log, environment
specification, and input provenance required for an independent party to
regenerate them. The criteria for success are that the plots agree with the
reference within a stated tolerance (or improve on it), that a second user can rerun the workflow
from the recorded artifacts alone, and that any changes from the reference
are reported. The last of these is the most demanding. The
validation chain in operation today exercises a fixed configuration maintained
by its authors, whereas the benchmark would present a geometry, sample, or
analysis setting not previously encountered, requiring the agent to identify
the appropriate configuration, detect a missing dependency, or recognize that a
plot has failed for physics rather than technical reasons. The distinction
between a job that has failed and one that has completed while producing an
incorrect result is where current agentic workflows are least reliable, and
where a physics-facing benchmark would be of greatest value.

\section{Verifiable AI workflows}

Across all of these subsystems, the same validation principles apply.

\begin{enumerate}[itemsep=1pt]
\item \textbf{Start from a physics decision.} Define what design or operational choice the model is meant to inform. A model without a decision context is difficult to validate.
\item \textbf{Choose metrics before training.} Metrics chosen after the model is built tend to reward what the model already does well.
\item \textbf{Validate tails and failures.} For \muc{} problems, losses, rare backgrounds, high occupancies, quenches, breakdown, and unusual signal signatures may dominate the risk.
\item \textbf{Track uncertainty and domain of validity.} Fast models are most useful when users know where they can and cannot be trusted.
\item \textbf{Preserve provenance.} Geometry, lattice, software version, random seeds, detector conditions, timing conventions, and intended use are part of the data product.
\item \textbf{Compare to trusted tools.} AI methods should be benchmarked against established simulations, test-stand data, or expert-reviewed reduced models.
\item \textbf{Make the benchmark rerunnable.} The community should be able to reproduce the validation plots without private knowledge.
\end{enumerate}

\section{Near-term steps}

The workshop was designed to establish connections among researchers from multiple R\&D areas, and actions were taken to encourage community engagement. Standing discussion channels were established for each major area: targetry, magnets and RF, ionization cooling, and the detector/software complex. The targetry discussion resulted in an agreement to coordinate with the existing high-power targetry programs at SNS, LBNF/DUNE, Mu2e, and J-PARC and the RaDIATE collaboration, with an integrated materials database as a first joint objective, aimed at benchmarking simulation against experiment.  In the ionization cooling discussion, the groups that have independently optimized individual cooling systems agreed to coordinate on coupled, global optimization, beginning with the capture and initial-cooling sections. In the detector discussion, the annual MODE workshops on differentiable programming for experiment design were identified as a natural venue for this work \cite{MODE:2026workshop}.  The importance of student engagement was emphasized.  Students are already engaged across the \muc{} effort---on the demonstrator target design, on cooling-channel optimization, on generative BIB---and it was agreed that continued student engagement is critical across all areas.

Broader community-organization efforts in HEP are converging on the practice of setting up grand challenges achievable on six-month timescales.  Building on that base, we recommend that the community organize around a small number of well-defined benchmark problems that demonstrate the advantages of AI workflows. Each grand challenge should have a physics owner, an AI/methods owner, a data owner, a validation owner, and a public workspace. Reproducibility, shared data and tools, working examples, and tutorials will be essential for broad adoption. Each challenge should be small enough for visible progress within months, but representative enough to make an impact. We identify four candidates, each with momentum already behind it:

\begin{enumerate}[itemsep=2pt]
\item \textbf{Cooling benchmark: differentiable or surrogate optimization of a restricted channel segment.} Define a common initial distribution, lattice segment, tunable parameters, loss definition, and validation suite. Compare manual, black-box, differentiable, and surrogate-assisted approaches. Require phase-space distributions, emittance evolution, transmission, tails, and parameter sensitivities. A differentiable cooling code whose first elements have been validated against established tracking through the IMCC benchmarking exercise provides a starting point.
\item \textbf{BIB benchmark: establish independent simulation samples and a validation suite before fast generation.} The immediate priority should be to reduce the cost of full BIB simulation and produce the largest feasible set of statistically independent events, together with an agreed detector geometry and validation suite. The benchmark should characterize spectra and occupancies as well as timing and spatial correlations, event-to-event fluctuations, rare high-occupancy tails, and reconstruction-level consequences. These samples should provide independent training and validation data for a subsequent generative-model challenge. Only once this foundation exists should conditional generators, including flows, diffusion, and other architectures, be compared on their ability to reproduce both detector-level distributions and the downstream reconstruction and design decisions obtained with full simulation.
\item \textbf{Detector/smart-pixel benchmark: local rejection under physics-safety constraints.} Define local detector information and hardware-inspired constraints. Compare simple cuts, interpretable classifiers, quantized ML, and neuromorphic approaches. Require performance on prompt and displaced benchmarks, building on the quantized-network results already in hand.
\item \textbf{Software-agent benchmark: standard validation report from a reference sample.} Build an agent-assisted but auditable workflow that creates a reproducible report from versioned inputs, exploiting the physics-validation machinery already running in the software stack's continuous integration. Measure success by repeatability, provenance, and agreement with reference plots.
\end{enumerate}

Given the limited size of the \muc{} community, these grand challenges should also be used as a mechanism for expanding the effective workforce. 

\section{Closing perspective}

The design of a \muc{} requires the scientific community to work at the boundary of accelerator physics, detector physics, superconducting technology, high-power targetry, software, and physics analysis. Typical design problems are coupled, high-dimensional, and expensive. As a result, it is expected that AI workflows can play a critical role. The opportunity of AI/ML is to change the tempo and scope of the design program: to move from isolated studies to validated, reusable, physics-aware design loops, and to do so with a community whose scarcest resource is expert time.

Success will depend on experts using these tools to ask sharper questions, explore larger design spaces, quantify uncertainty, and make their work reproducible. The resulting workflows should also lower barriers for the students and new collaborators the ``muon shot'' will need.

\section*{Acknowledgments}

We thank the speakers and participants of the AI4MuC workshop, and all members of the high energy physics, nuclear physics, and AI/ML communities who participated in the creation of this report. 

Generative AI was used in the production of this report: the live transcripts, slides, and meeting notes were used to create a first draft. All sections were then edited by the report authors. The authors are fully responsible for the content of the report.

\bibliographystyle{unsrt}
\bibliography{apssamp}

\end{document}